\documentclass[twocolumn,prl,floatfix,citeautoscript,nofootinbib,superscriptaddress,amsmath,amssymb]{revtex4-2}
\usepackage{graphicx}% Include figure files
\usepackage{dcolumn}% Align table columns on decimal point
\usepackage{bm}% bold math
\usepackage{color}
\usepackage{braket}
\usepackage{float}
\usepackage{adjustbox}
\usepackage[caption=false]{subfig}
\usepackage{nicefrac}
\usepackage{soul}
\usepackage{amsmath}
\usepackage{mathtools}

\usepackage{dcolumn}% Align table columns on decimal point
\newcommand{\kbar}{\mathchar'26\mkern-9mu k}
\newcommand{\Ez}{\langle E_z \rangle}
\newcommand{\Erec}{E_{\rm rec}}
\newcommand{\gn}{gn_{\rm 1D}}

\usepackage{comment}
\usepackage{hyperref}
\hypersetup{
	colorlinks   = true, %Colours links instead of ugly boxes
	urlcolor     = blue, %Colour for external hyperlinks
	linkcolor    = blue, %Colour of internal links
	citecolor   = blue %Colour of citations
}
\begin{document}

\title{Many-Body Anderson Metal-Insulator Transition using Kicked Quantum Gases}

\author{Jun Hui See Toh}
\affiliation{Department of Physics, University of Washington, Seattle, WA, USA}
\author{Mengxin Du}
\affiliation{Department of Physics, The University of Texas at Dallas, Richardson, TX, USA}
\author{Xinxin Tang}
\affiliation{Department of Physics, University of Washington, Seattle, WA, USA}
\author{Ying Su}
\affiliation{Department of Physics, The University of Texas at Dallas, Richardson, TX, USA}
\author{Tristan Rojo}
\affiliation{Department of Physics, University of Washington, Seattle, WA, USA}
\author{Carson O. Patterson}
\affiliation{Department of Physics, University of Washington, Seattle, WA, USA}
\author{Nicolas R. Williams}
\affiliation{Department of Physics, University of Washington, Seattle, WA, USA}
\author{Chuanwei Zhang}
\email{chuanwei.zhang@utdallas.edu}
\affiliation{Department of Physics, The University of Texas at Dallas, Richardson, TX, USA}
\author{Subhadeep Gupta}
\email{deepg@uw.edu}
\affiliation{Department of Physics, University of Washington, Seattle, WA, USA}

\date{\today}

\begin{abstract}
Understanding the interplay of interactions and disorder in quantum transport poses long-standing scientific challenges, with many-body quantum transport phenomena in high-dimensional disordered systems remaining largely unexplored experimentally. We utilize a momentum space lattice platform using quasi-periodically kicked ultracold atomic gases to experimentally investigate many-body effects on the three-dimensional Anderson metal-insulator transition. We observe interaction-driven sub-diffusion and a divergence of delocalization onset time on approaching the many-body phase boundary. Mean-field numerical simulations are in qualitative agreement with experimental observations.

\end{abstract}

\maketitle

Conceived more than sixty years ago, the celebrated Anderson picture of electron transport in the presence of disorder \cite{ande58,abra79} predicts localization in one and two dimensions ($d=1,2$), and delocalization in $d>2$ below a critical minimum disorder, demarcating a metal-insulator transition (MIT). How inter-particle interactions compete with disorder during quantum transport has since been the subject of intense theoretical scrutiny \cite{flei80,natt03,gorn05,bask06} with to-date unresolved questions surrounding the fate of localized states in the presence of interactions. Landmark experiments within the last fifteen years with ultracold atoms in spatially disordered optical lattices have observed interaction-driven transport phenomena, such as sub-diffusive delocalization \cite{deis10,luci11}, many-body localization \cite{schr15,lusc17,luki19,kond15}, and a
finite temperature Bose glass-superfluid transition \cite{meld16}. However, the effects of many-body interactions on quantum transport phenomena in high-dimensional disordered systems, such as the $d=3$ Anderson MIT \cite{chab08,kond11,jend12}, have largely remained unexplored experimentally.

In recent years, lattices in the synthetic dimension of momentum space have become a fertile alternate avenue for dynamical studies and quantum simulation with ultracold atoms \cite{meie18}. The $d=1$ Anderson Hamiltonian can be simulated in momentum space using the quantum kicked rotor, where ultracold atoms are periodically driven or ``kicked" by a pulsed standing wave and can exhibit the corresponding ``dynamical" localization \cite {moor94, moor95, dArc01, amma98, wimb03, Duffy2004, ulla12, gadw13}. Very recently, experiments have observed the interaction-driven breakdown of this dynamical localization \cite{seet22, caoa22}. By modulating the strength of the kicks, $d$-dimensional Anderson models can also be engineered in the quasi-periodic kicked rotor (QPKR), with the pseudo-random phase accumulated by atoms at different momenta corresponding to the disorder, kick strength to inter-site tunneling, and number of modulation frequencies to $d-1$ \cite{casati89, lema09, fish82, grem84}. The QPKR technique has been utilized experimentally to simulate disordered non-interacting systems in $d=2$ \cite{mana15} and $d=3$, where the Anderson metal-insulator transition was also observed \cite{chab08}. Recent theoretical works incorporating mean-field interactions predict sub-diffusive delocalization in the $d=3$ Anderson insulator region but disagree on the value of the corresponding sub-diffusive exponent \cite{cher14,erma14,verm20}.

In this work we use the QPKR in conjunction with atomic interaction tuning to experimentally observe interaction-driven delocalization of the Anderson insulator and the many-body Anderson MIT in $d=3$. The transition boundary is manifest as a divergence of delocalization onset time with varying kick strength. We study the inverse relation of onset time with interaction strength and also demonstrate interaction-driven delocalization in $d=2$ and $d=4$ synthetic space. Our numerical mean-field simulations incorporating system inhomogeneity capture the general trends of our observations, but with quantitative deviations that grow with increasing interaction strength. 

\begin{figure}[h!]
		\center
		\includegraphics[width=0.46\textwidth]{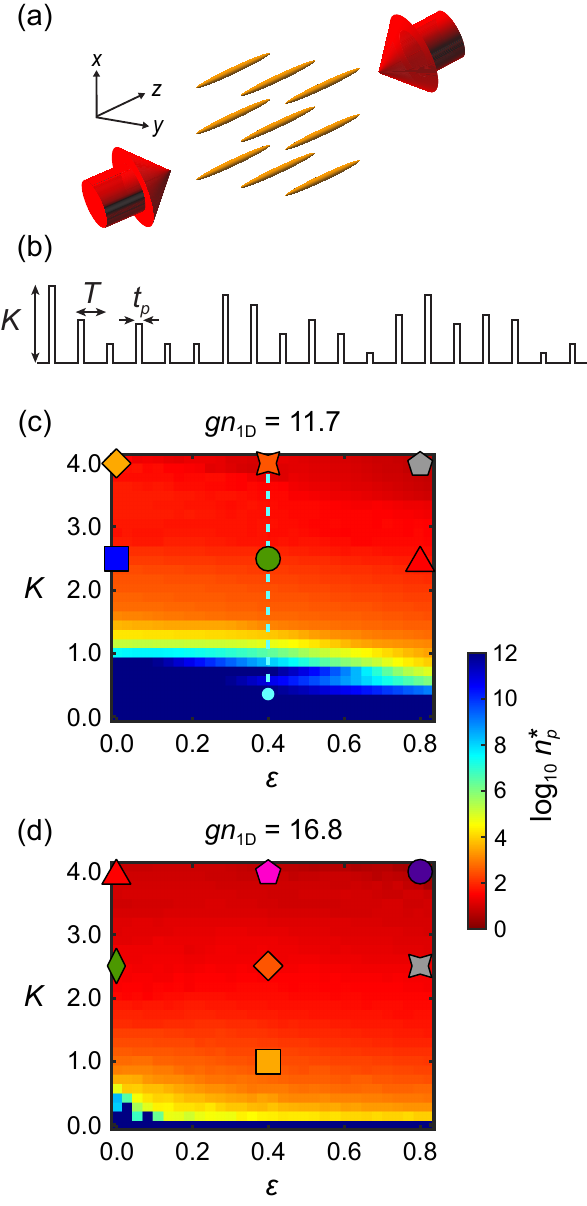}
		\caption{Experimental scheme and phase diagrams for many-body Anderson model with different interaction strengths. (a) Experimental schematic showing BECs in 1D tubes with kicking pulses applied along the axial ($z$) direction. (b) Schematic of quasi-periodic drive of optical standing wave pulses to engineer the $d=3$ Anderson model. In (c) and (d) we show the numerically simulated phase diagrams of the interacting $d=3$ Anderson model using the Gross-Pitaevskii equation for two different interaction strengths $gn_{\rm 1D}=11.7$ and $16.8$. The sharp change in the delocalization onset time $n_p^*$ marks the Anderson MIT. Markers indicate where $d=3$ delocalization data was collected (Figs. \ref{fig:qpkr_fig2}, \ref{fig:qpkr_fig3}, \ref{fig:qpkr_fig4}) and the dashed line indicates the line along which the many-body Anderson transition was observed (Fig. \ref{fig:qpkr_fig2}).}
		\label{fig:qpkr_fig1}
\end{figure}

\begin{figure*}[ht!]
		\center
		\includegraphics[width=1\textwidth]{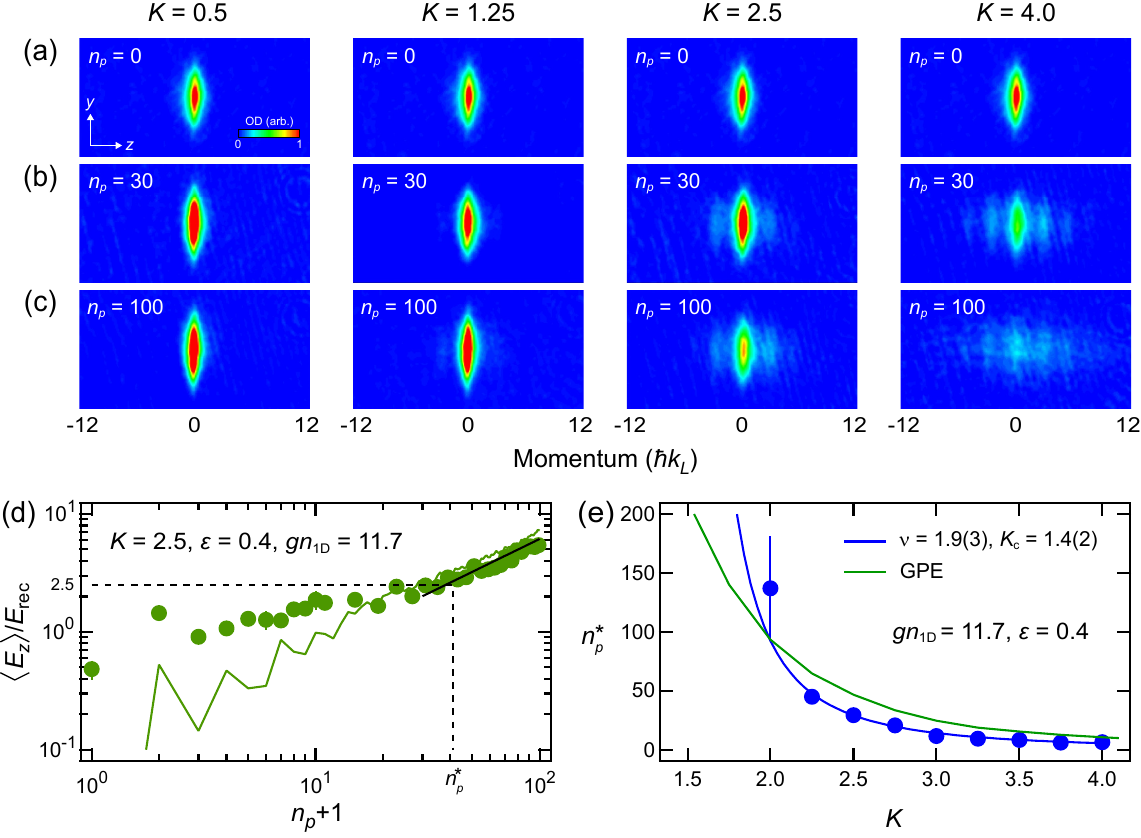}
		\caption{Observing the many-body Anderson MIT in three-dimensional synthetic momentum space. (a-c) Sequence of absorption images for $n_p=0,30,100$ showing the transition from localized to delocalized evolution as $K$ is varied from 0.5-4.0 for $\gn=11.7$ and $\varepsilon = 0.4$. The time-of-flight for all images is $15\,$ms. The kicks are applied along the horizontal ($z$) direction coinciding with the axis of the 1D gas. (d) Evolution of axial kinetic energy corresponding to $K=2.5$. The solid line is a power law fit with an exponent of 0.70(2). The dashed lines mark $E^*$ and $n_p^*$. (e) Divergence of $n_p^*$ as $K$ approaches a critical value, together with an inverse power-law fit (blue line). The green line shows the mean-field theory result.}
		\label{fig:qpkr_fig2}
\end{figure*}

Our experimental approach \cite{SUPP} for quantum simulation of the many-body Anderson model utilizes Bose-Einstein condensates in the quasi-1D regime which are kicked along the axial ($z$) direction using a pulsed optical standing wave lattice with period $T$ and pulse width $t_p \ll T$ (see Fig. \ref{fig:qpkr_fig1}(a)). $1/T$ is incommensurate with the recoil frequency $\omega_{\rm rec}=\hbar k^2_L/ 2m = E_{\rm rec}/\hbar$, where $k_L = 2\pi/\lambda$ is the wavevector and $m$ is the atom mass. We engineer the $d$-dimensional Anderson model (for $d=1$ to 4) in momentum space by modulating the amplitude of these kick pulses with $d-1$ additional incommensurate frequencies.

A mean-field description of the many-body dynamics of the $d=3$ system is captured by the following dimensionless non-linear Gross-Pitaevskii (GP) equation
\begin{multline}
i\kbar \partial_\tau \Phi(\theta,\tau) =  \Bigg( \Bigg. -\frac{\kbar^2}{2} \partial^2_\theta + \frac{1}{2}{\omega}_\theta^2\theta^2  + g |\Phi(\theta,\tau)|^2\\- K (1 +  \varepsilon \cos \omega_2 \tau \cos \omega_3 \tau) \cos \theta \sum_{n_p} \delta(\tau-n_p)  \Bigg) \Bigg. \Phi(\theta,\tau)
\label{eq:GP}
\end{multline}
where $\Phi$ is the wavefunction, $\kbar = 8\omega_{\rm rec} T$ is the dimensionless reduced Planck constant, $\theta = 2k_L z$ and $\tau=t/T$ are the dimensionless position and time parameters, and $n_p$ is the pulse number. The dimensionless axial frequency is ${\omega}_\theta=\omega_zT$ and the dimensionless initial peak density is $n_\text{1D}=|\Phi(0,0)|^2 = {\bar n}_{\rm 1D}/2k_L$, where the wavefunction is normalized as $\int d\theta |\Phi(\theta,\tau)|^2 =N_\text{atom}$. The dimensionless kick strength $K$ and interaction constant $g$ are defined as
\begin{equation}
    K = 4s_z \omega_{\rm rec}^2 t_p T \, ,\quad g=\frac{2\bar{g} k_L \kbar T}{\hbar} = \kbar^2 \frac{k_L a_s}{(k_L a_\perp)^2} \, ,
    \label{eq:params}
\end{equation}
where $s_z E_{\rm rec}$ is the peak depth of the pulsed optical lattice. $K$ is modulated by incommensurate frequencies $\omega_2$ and $\omega_3$ (in units of $1/T$) with modulation strength $\varepsilon$. Throughout this work, we use $\omega_2=2\pi\times \sqrt{2}$ and $\omega_3=2\pi\times \sqrt{3}$ (see Fig. \ref{fig:qpkr_fig1}(b)), with $t_p=4\,\mu s$, $T=105\,\mu s$ and $\kbar=5.26$, for our QPKR realization of the $d=3$ Anderson model. The interaction strength $gn_{\rm 1D}$ is controlled experimentally through $a_{\perp}$, the transverse oscillator length of confinement \cite{SUPP}.

\begin{figure}[h!]
		\center
		\includegraphics[width=0.45\textwidth]{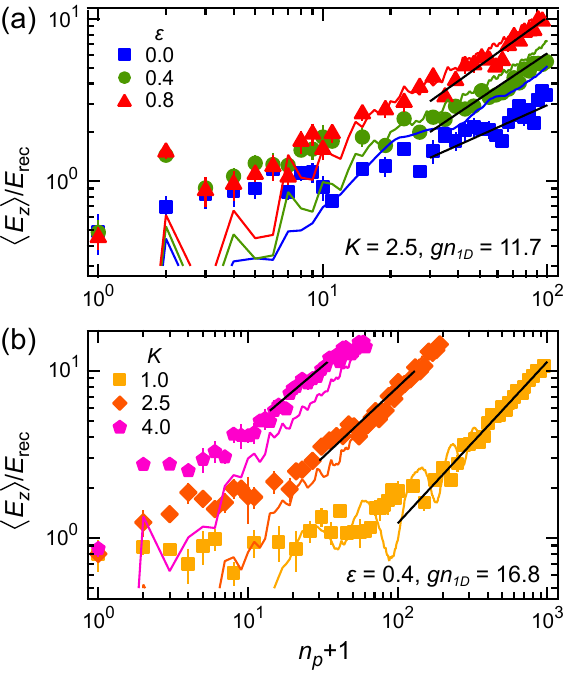}
		\caption{Evolution of axial kinetic energy $\langle E_z \rangle$ with pulse number $n_p$ for different (a) modulation strengths $\varepsilon$ and (b) kick strengths $K$. The colored solid lines are the corresponding numerical simulations using the GP equation. The black solid lines are power law fits to the data.}
		\label{fig:qpkr_fig3}
\end{figure}

In Figs. \ref{fig:qpkr_fig1}(c) and (d), we show the numerically-simulated $K-\varepsilon$ phase diagram for two different interaction strengths, where the transition from a localized insulating phase to a delocalized metallic phase can be clearly identified across a sharp phase boundary \cite{SUPP}. To remove unimportant single-particle features arising from the harmonic oscillator timescale and the finite depth trap \cite{bill09, SUPP}, and more clearly reveal the effects of interactions, we display the phase diagrams through a delocalization onset time rather than diffusive exponent. Importantly, the interaction-shifted Anderson MIT boundaries are in agreement for either parameter choice \cite{SUPP}.  

We define the onset time $n_p^*$ as the pulse number at which the average energy $\langle E_z\rangle$ of a system undergoing delocalization reaches $E^*$, chosen to be much larger than the initial energy. For an evolution governed by $\langle E_z\rangle=E_0 (n_p)^{\alpha}$ with diffusive exponent $\alpha$, $n_p^* = (E^*/E_0)^{1/\alpha}$ is directly related to the inverse of the diffusion constant $E_0$, whose divergent behavior is used to characterize the MIT phase boundary on the metal side in the homogeneous non-interacting case \cite{lema09}. In our inhomogeneous system, we use $n_p^*$ for this characterization, with the insulator phase corresponding to $n_p^* \rightarrow \infty$. The insulating regions of these phase diagrams are far smaller than that of the non-interacting homogeneous Anderson model \cite{chab08, lope13}, with the higher interaction strength showing a smaller insulating region. 

We present our observation of a many-body Anderson MIT in Fig. \ref{fig:qpkr_fig2} where sequences of absorption images for $n_p = 0, 30, 100$ (Fig. \ref{fig:qpkr_fig2}(a,b,c)) show the growth in the axial ($z$) momentum distribution for different kick strengths $K$, with fixed interaction strength $gn_{\rm 1D}=11.7$ and $\varepsilon=0.4$. A clear progression from insulating (localized) to metal (delocalized) behavior emerges as the $K$ parameter is increased. As shown in Fig. \ref{fig:qpkr_fig2}(d), the evolution of the axial energy $\langle E_z\rangle$ on the delocalized side exhibits some initial coherent single-particle dynamics before increasing monotonically for larger pulse numbers. Fitting a power law to this long time evolution $\langle E_z\rangle=E_0 (n_p)^{\alpha}$ allows extraction of $\alpha$ and $E_0$, and therefore $n_p^*$ for a chosen $E^*$. $\alpha$ and $E_0$ are the slope and intercept on the log-log plot of $\langle E_z\rangle$ vs $n_p$, respectively, and can also be obtained from the $\langle E_z\rangle$ values at two different $n_p$ \cite{SUPP}. Throughout this paper we define $n_p^*$ using $E^*=2.5E_{\rm rec}$, a sufficiently large value which also provides an adequate dynamic range for data analysis.   

We observe a divergence of $n_p^*$ as $K$ is varied with $gn_{\rm 1D}=11.7$ and $\varepsilon=0.4$ (Fig. \ref{fig:qpkr_fig2}(e)) which marks the phase transition boundary. Fitting to the power law $1/|(K-K_c)|^\nu$ yields $\nu=1.9(3)$ and $K_c=1.4(2)$. This value of $\nu$ is independent (within error bars) of choice of $E^*$ \cite{SUPP}. In comparison, $\nu = 1.58$ for the non-interacting Anderson model \cite{cher14,verm20}. Our numerical mean-field simulations capture the general trend of the experimental observations, as shown by the green solid line in Fig. \ref{fig:qpkr_fig2}(f) and also seen in Fig. \ref{fig:qpkr_fig1}(c). 

In Fig. \ref{fig:qpkr_fig3}, we show the evolution of the axial kinetic energy of the system for a few different combinations of $\varepsilon$, $K$, and $gn_{\rm 1D}$, corresponding to various locations in the phase diagrams of Fig. \ref{fig:qpkr_fig1}. As shown in Fig. \ref{fig:qpkr_fig3}(a), the interacting system delocalizes, exhibiting sub-diffusive dynamics and an earlier delocalization onset with stronger modulation strength. This can be understood by noting that as $\varepsilon \rightarrow 0$, the system approaches $d=1$, and delocalization is impeded by the reduced dimensionality, similar to the well-known non-interacting situation where a $d=1$ system with arbitrarily weak disorder is always localized. Varying the kick strength $K$ also results in a similar but stronger response as shown in Fig. \ref{fig:qpkr_fig3}(b), where the onset time shortens by more than one order of magnitude as $K$ is varied from 1 to 4. This trend is expected as $K$ is the momentum-space analog of the tunneling energy in the Anderson model.   

Our mean-field simulations show reasonable agreement with the experimental observations. Power law fits return exponents $\alpha \in \{0.6,1.1\}$, and are predominantly sub-diffusive. These values characterize all our observations shown in the main text and also additional data included in the supplemental materials, and are larger than mean-field theory predictions for a homogeneous system \cite{cher14,erma14}. We note that small regions with $\alpha > 1$ also exist in the parameter space explored in our inhomogeneous system \cite{SUPP}. 

\begin{figure}[h!]
		\center
		\includegraphics[width=0.45\textwidth]{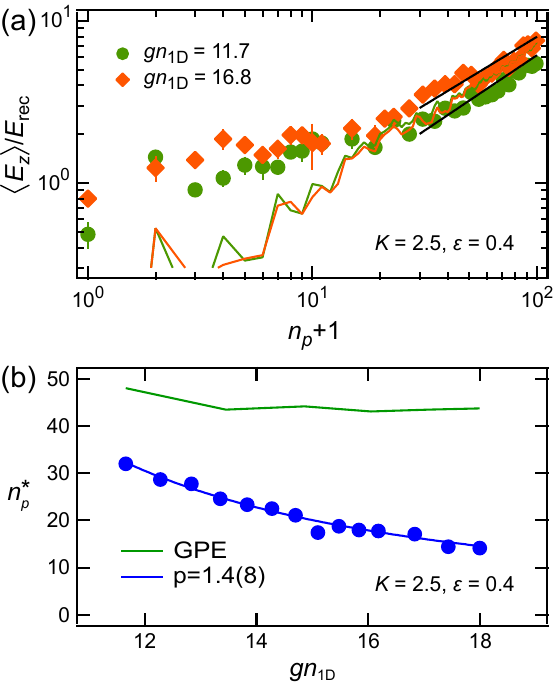}
		\caption{Tuning interaction strength in the $d=3$ Anderson model. (a) Evolution of axial kinetic energy $\langle E_z \rangle$ with pulse number $n_p$ for different interaction strengths $\gn$. The colored solid lines are the corresponding numerical simulations using the GP equation. The black solid lines are power law fits to the data. (b) Onset of delocalization $n_p^*$ as a function of interaction strength $\gn$. The green solid line shows the numerical simulation using the GP equation. The blue solid line shows an inverse power-law fit.}
		\label{fig:qpkr_fig4}
\end{figure}

To more systematically analyze the interaction dependence of the system, we fix kick and modulation strengths to representative values $K=2.5$ and $\varepsilon=0.4$, and study the evolution of axial energy with kick number across the range of experimentally available interaction strengths. The time evolution of both $gn_{\rm 1D}=11.7$ and $16.8$ cases in Fig. \ref{fig:qpkr_fig4}(a) exhibit delocalization with a small difference in character. 

The interaction dependence can be more sensitively detected using onset time $n_p^*$ as the metric, rather than long-time energy diffusion. In Fig. \ref{fig:qpkr_fig4}(b) we show how $n_p^*$ changes as we vary the interaction strength by tuning the external confinement and thus $a_\perp$. Consistent with the phase diagrams in Fig. \ref{fig:qpkr_fig1}, we observe that delocalization onset time is earlier with stronger interaction strength. Fitting an inverse power law of the form $1/(gn_{\rm 1D}-gn_{\rm 1D,c})^p$ to the data returns a value of $p=1.4(8)$ for a range of $E^*$ \cite{SUPP}. The deviation from our mean-field numerical model (green line in Fig. \ref{fig:qpkr_fig4}(b)) is greater for larger interaction strengths. 

By experimentally realizing an interacting QPKR Hamiltonian simulator, we investigated interaction effects on the Anderson metal-insulator transition. A many-body $d=3$ Anderson MIT was observed as a divergence in the delocalization onset time, a measure of the inverse diffusion rate, with a critical exponent of 1.9(3). Measured sub-diffusive delocalization exponents in the metal phase exceed prior mean-field theory predictions for the homogeneous case. Incorporating experimental inhomogeneity into the mean-field theory, we find qualitative agreement with our observations with deviations increasing with interaction strength. Using the flexibility of the momentum-space approach, we have also engineered Anderson models in other dimensions ($d=1-4$) and observed many-body sub-diffusive delocalization of the insulator phase with shorter onset time at larger interaction strength \cite{SUPP}. Our combined experiment-theory collaboration advances the area of many-body quantum simulation of high-dimensional quantum transport, while revealing the need for a beyond mean-field theoretical description to address the infinitely long-range interaction in momentum space \cite{seet22}. 

Future work includes experimental realization of the interacting QPKR in a homogeneous 1D system, such as a ring trap \cite{rama11b,ecke18,caiy22,delp22} or box trap \cite{gaun13}, where potential universality \cite{cher14} of the sub-diffusive exponent $\alpha$ can be tested and the fate of the many-body Anderson MIT can be examined in a homogeneous setting. Further studies of the nature of many-body transport phases can be pursued with a larger interaction tuning range using a magnetic Feshbach resonance \cite{chin10,caoa22}. The technique of kicked quantum gases is well suited for studying the interplay of disorder, interactions and dimensionality in quantum transport, and fruitful applications to other ultracold many-body systems such as the Tonks gas \cite{vuat21} and strongly-interacting fermions \cite{mess18} are envisioned. 

\begin{acknowledgments}
\textbf{Acknowledgements}: We thank M. Rudner, C.R. Laumann, and A. Chandran for helpful discussions. Work at the University of Washington is supported by the Air Force Office of Scientific Research (FA9550-22-1-0240). C.O.P. acknowledges support from NSF-NRT grant Number DGE-2021540. T.R. acknowledges support from the NSF-REU program at the University of Washington (Grant No NSF-REU 1851741). Work at the University of Texas at Dallas is supported by the Air Force Office of Scientific Research
(FA9550-20-1-0220), Army Research Office (W911NF-17-1-0128), Department of Energy (DESC0022069) and National Science Foundation (PHY-2110212, OMR-2228725). 
\end{acknowledgments}

\clearpage

\onecolumngrid

\renewcommand{\theequation}{S\arabic{equation}}
\renewcommand{\thefigure}{S\arabic{figure}}
\setcounter{equation}{0}
\setcounter{figure}{0}

%\usepackage{amsbsy}
%\usepackage{latexsym,epsfig,graphicx}
%\usepackage{dcolumn}
%\usepackage{graphicx}
%\usepackage{subfigure}
%\usepackage{comment}
%\usepackage{color}
%\usepackage{bm}
%\usepackage{mathrsfs}
%\usepackage{amsfonts}
%\usepackage{amsmath}
%\usepackage{amssymb}
%\usepackage{xspace}
%\usepackage{epstopdf}
%\usepackage{tabularx}
%\usepackage{longtable}
%\usepackage[colorlinks=true, letterpaper=true, pdfstartview=FitV, linkcolor=blue, citecolor=blue, urlcolor=blue]{hyperref}
%\usepackage[normalem]{ulem}
%\usepackage{soul}  %for highlighting text
%\usepackage{rotating}
%\usepackage{lineno}
%\linenumbers

%\usepackage{xr}
\makeatletter
\newcommand*{\addFileDependency}[1]{% argument=file name and extension
\typeout{(#1)}% latexmk will find this if $recorder=0
% however, in that case, it will ignore #1 if it is a .aux or 
% .pdf file etc and it exists! If it doesn't exist, it will appear 
% in the list of dependents regardless)
%
% Write the following if you want it to appear in \listfiles 
% --- although not really necessary and latexmk doesn't use this
%
\@addtofilelist{#1}
%
% latexmk will find this message if #1 doesn't exist (yet)
\IfFileExists{#1}{}{\typeout{No file #1.}}
}\makeatother

\newcommand*{\myexternaldocument}[1]{%
\externaldocument{#1}%
\addFileDependency{#1.tex}%
\addFileDependency{#1.aux}%
}

%\myexternaldocument{main}

\renewcommand{\theequation}{S\arabic{equation}}
\renewcommand{\thefigure}{S\arabic{figure}}

\setcounter{MaxMatrixCols}{10}

\pdfoutput=1

\begin{center}

{\bf \large Supplemental Material for:\\ 
Many-Body Anderson Metal-Insulator Transition using Kicked Quantum Gases}
\end{center}

\section {Experimental Setup and Procedures}

The experimental setup is similar to earlier work \cite{seet22} and we will summarize it briefly here. We first prepare a BEC of $^{174}$Yb with $1.5 \times 10^5$ atoms in a crossed-beam optical dipole trap (ODT) with trapping frequencies $\{\omega_{0x}, \omega_{0y}, \omega_{0z}\}=2\pi \times \{145, 16, 53\}$ Hz and chemical potential $h \times 1.1$ kHz. The BEC is then adiabatically loaded into an array of 1D tubes formed by a two-dimensional optical lattice of wavelength $\lambda=1073$ nm with beam waists $\{w_x, w_y\}=\{121, 101\}\,\mu$m. The potential depth $V_{\rm lat}$ of each arm of the lattice can be expressed as a unitless parameter $s_{\perp}=V_{\rm lat}/E_{\rm rec}$, where $E_{\rm rec}=\hbar \omega_{\rm rec}$ is the recoil energy. For a typical lattice depth $s_{\perp}=90$ used in this work, the transverse and axial trapping frequencies (proportional to $\sqrt{s_{\perp}}$) of the tubes are $\omega_{\perp}= 2\pi \times 19$ kHz and $\omega_z = 2\pi \times 59$ Hz respectively. We estimate a peak particle number $N_{\rm atom}=650$ and an initial 1D peak density of $\bar{n}_{\rm 1D} = 24/\,\mu$m for the central tube. Our system is quasi-1D such that the gas is kinematically 1D ($\mu, k_B T \ll \hbar \omega_\perp$) with the two-body scattering length $a_s=5.55\,$nm retaining its 3D value. The starting BEC fraction in the tubes is higher than $85\%$, and intertube tunneling is negligible. Higher lattice depth $s_{\perp}$ results in a shorter transverse oscillator length $a_{\perp} \propto s_{\perp}^{1/4}$ in the tubes and thus higher 1D density. Since the atom number varies across different 1D tubes, we quote a weighted average of the interaction strength throughout this paper. The measured background heating rate (in the absence of pulses) is $6 E_{\rm rec}$/s for the highest lattice depths used, which is negligible on the timescale of all the reported experiments.

The kicks are applied to the atoms along the axial direction of the 1D tubes using another optical lattice, with wavelength $\lambda=1073$ nm and beam waist $w_z=99\,\mu$m, pulsing at kick period $T$ and pulse width $t_p$. The strong confinement in the 1D tubes $\omega_\perp \gg \omega_{\rm rec}$, suppresses excitation of the transverse modes, allowing the energy from the pulses to transfer fully to the axial dynamics of the system. After the desired pulse number $n_p$, we diabatically turn off all traps and take a time-of-flight absorption image. The axial momentum distribution is obtained by integrating the absorption
image along $y$. The axial kinetic energy of the atoms $\langle E_z \rangle$ is then calculated from the observed axial momentum distribution.

\section{Interacting Anderson Model in different synthetic dimensions}

\begin{figure}[h!]
		\center
		\includegraphics[width=0.92\textwidth]{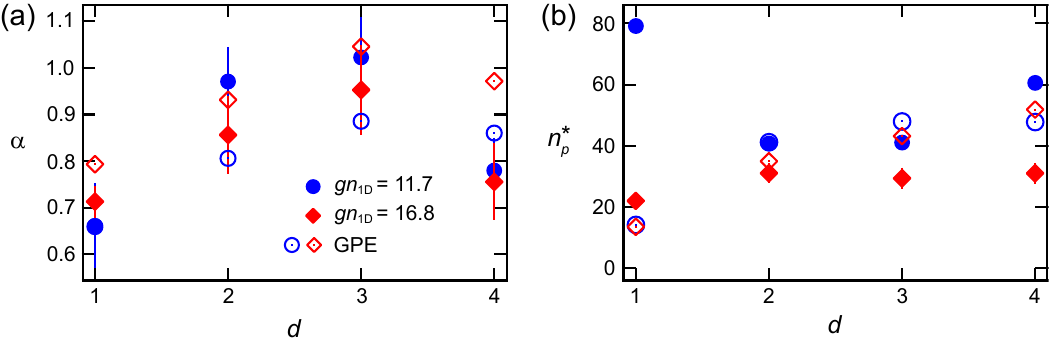}
		\caption{Interaction-driven delocalization in $d$-dimensional Anderson models in the synthetic momentum space. (a) Diffusion exponent $\alpha$ and (b) onset time of delocalization $n_p^*$ of $d$-dimensional systems with $d=1-4$ for two different interaction strengths $gn_{\rm 1D}=11.7$ and 16.8. The filled markers are the experimental data and the empty markers are numerical simulations using the GP equation. For the $d=2$ cases, $\omega_2=2\pi \times \sqrt{5}$. For the $d=3$ cases, $\omega_2=2\pi \times \sqrt{2}$, $\omega_3=2\pi \times \sqrt{3}$. For the $d=4$ cases, $\omega_2=2\pi \times \sqrt{3}$, $\omega_3=2\pi \times \sqrt{5}$, $\omega_4=2\pi \times \sqrt{7}$. For all cases, $K=2.5$ and $\varepsilon=0.4$.}
		\label{fig:qpkr_fig5}
\end{figure}

We use the flexibility of the QPKR to engineer arbitrary-dimensional Anderson models in the synthetic momentum space and explore the dependence of delocalization behavior on dimension number $d$. In Fig. \ref{fig:qpkr_fig5} we show the extracted $\alpha$ and $n_p^*$ from energy evolution curves in different dimensions $d=1-4$, implemented using $ \varepsilon \prod_{i=0}^{d-1} \cos(\omega_i \tau)$ for the modulation term in Eqn.(1) of the main text, with incommensurate $\omega_i$ and keeping $K=2.5$ and $\varepsilon=0.4$. The experimental sub-diffusive exponent $\alpha$ for $d=1-4$ and both interaction strengths (Fig. \ref{fig:qpkr_fig5}(a)) all lie between 0.6 and 1. These values are in reasonable agreement with our numerical simulations, while being generally higher than earlier calculations performed in the homogeneous case \cite{cher14,erma14}. The onset times (Fig. \ref{fig:qpkr_fig5}(b)) show the expected trend of being shorter for the stronger interaction case. There is also reasonable agreement between the experimental and numerical values, except for the $d=1$, $gn_{\rm 1D}=11.7$ case. We note that this falls close to the localization-delocalization boundary investigated in earlier work \cite{seet22}, where strong deviations between experiment and mean field theory were reported.

\section{Experimental Data Analysis Details}

\begin{figure}[h!]
 
\begin{center}
\begin{tabular}{|c|c|c|c|} 
\hline
        & $\varepsilon=0.0$ &  $\varepsilon=0.4$ &  $\varepsilon=0.8$   \\ 
\hline
\rotatebox{90}{\qquad \qquad \quad $K=2.5$} & \includegraphics[width=0.3\textwidth]{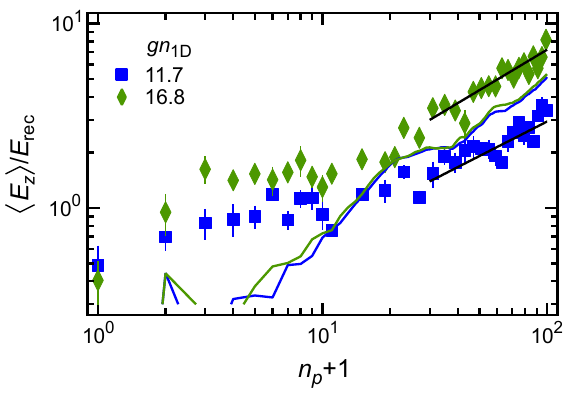} & \includegraphics[width=0.3\textwidth]{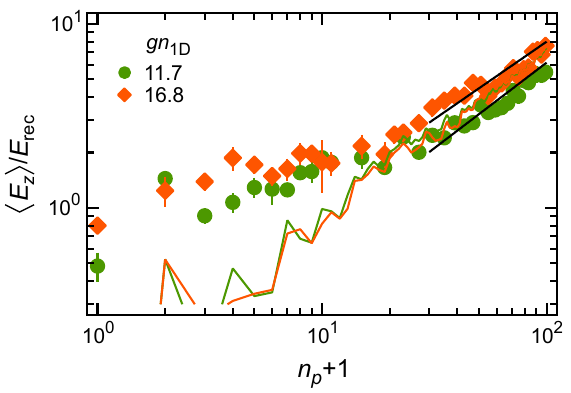} & \includegraphics[width=0.3\textwidth]{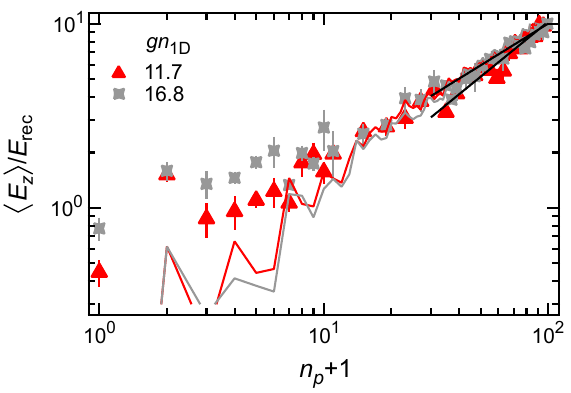}    \\ 
\hline
\rotatebox{90}{\qquad \qquad \quad $K=4.0$} & \includegraphics[width=0.3\textwidth]{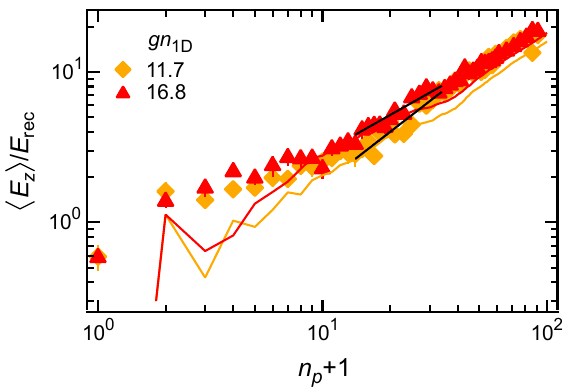} & \includegraphics[width=0.3\textwidth]{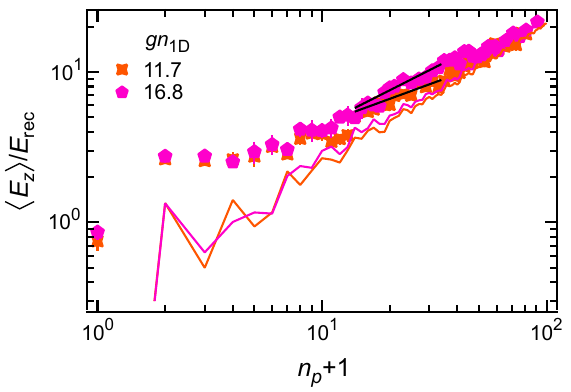} & \includegraphics[width=0.3\textwidth]{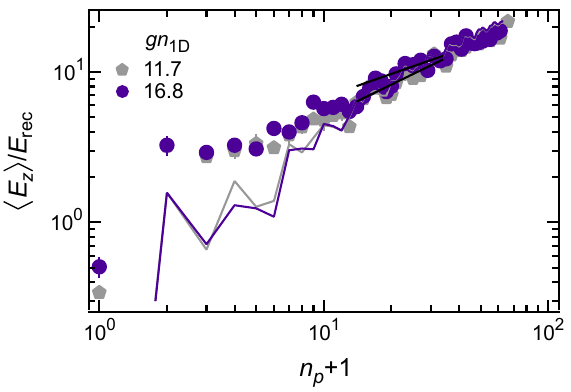}     \\ 
\hline
\end{tabular}
\end{center}

\caption{Energy Evolution Data for $d=3$ with different $K,\varepsilon,gn_{\rm 1D}$, together with corresponding GP numerical simulation and power law fit.}
\label{fig:qpkr_figS1}
\end{figure}

\begin{figure}[h!]
\begin{center}
\begin{tabular}{|c|c|c|} 
\hline
 $d=1$ &  $d=2$ &  $d=4$   \\ 
\hline
\includegraphics[width=0.3\textwidth]{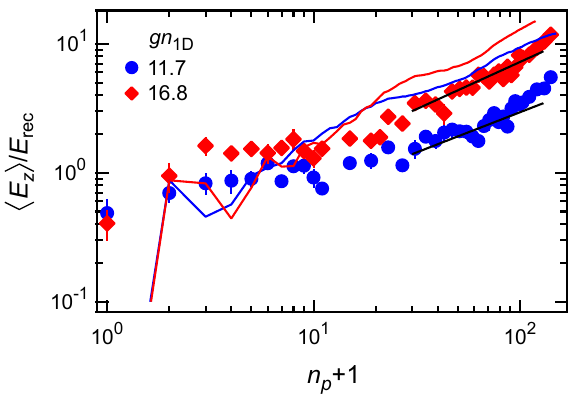} & \includegraphics[width=0.3\textwidth]{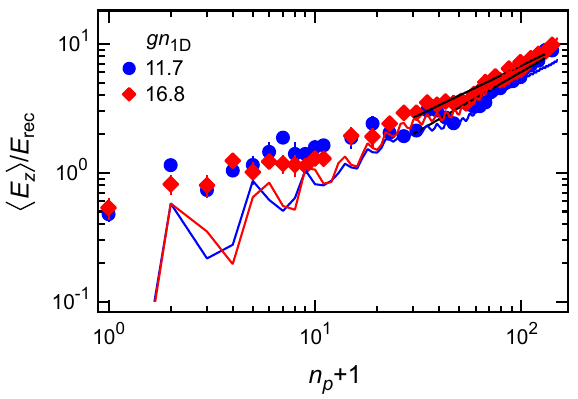} & \includegraphics[width=0.3\textwidth]{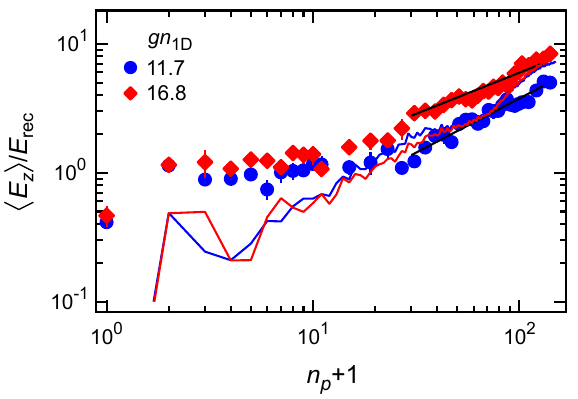}     \\ 
\hline
\end{tabular}
\end{center}
\caption{Energy Evolution Data for $d=1,2,4$ with $gn_{\rm 1D}=$11.7 and 16.8, $K=2.5$ and $\varepsilon=0.4$, together with corresponding GP numerical simulation and power law fit.}
\label{fig:qpkr_figS2}
\end{figure}

\noindent {\it Power Law Fitting for Energy Evolution Data.---} In Fig. \ref{fig:qpkr_figS1} we show the energy evolution datasets collected for various values of $K$, $\varepsilon$, and $gn_{\rm 1D}$ for $d=3$. The graphs also show the corresponding GP numerical simulation and the power law fit $E_0 n_p^{\alpha}$. The fitting ranges are indicated, together with the total number of datapoints in Table \ref{tab:qpkr_tab1}. The fitted diffusion constant $E_0$, exponent $\alpha$, and onset time $n_p^*$ are also listed therein. Fig. \ref{fig:qpkr_figS2} and Table \ref{tab:qpkr_tab2} show the corresponding data and analysis details for the $d=1,2,4$ cases. 

\begin{center}

\begin{table}[h]
\caption{\bf{Ranges of pulse numbers used for fitting the data in Fig. \ref{fig:qpkr_figS1}, corresponding energy ranges, number of fits, and extracted values.}}
\begin{tabular}{|c|c|c|c|c|c|c|c|c|c|c|c|} 
\hline
Dataset  & $K$ & $gn_{\rm 1D}$ & $\varepsilon$ & Beginning $n_p$ & Ending $n_p$  & Beginning $\Ez/\Erec$ & Ending $\Ez/\Erec$ & Number of fits & $E_0$ & $\alpha$ & $n_p^*$ \\
\hline
(1) & 1.0 & 16.8 & 0.4 & $100-300$ & $800-1000$ & $1.6 - 3.6 $ & $9.3 - 10.7 $  & 45 & 0.013(2) & 0.98(3) & 212(7)  \\ 
(2) & 2.5 & 11.7 & 0.0 & $30-50$   & $110-130$  & $1.5 - 2.2 $ & $3.9 - 4.5 $   & 18 & 0.16(7)  & 0.7(1)  & 79(1)   \\ 
(3) & 2.5 & 16.8 & 0.0 & $30-50$   & $110-130$  & $2.9 - 4.5 $ & $8.2 - 10.4 $  & 18 & 0.27(4)  & 0.71(3) & 23(1)   \\ 
(4) & 2.5 & 11.7 & 0.4 & $30-50$   & $110-130$  & $2.4 - 3.6 $ & $7.6 - 8.7 $   & 18 & 0.06(2)  & 1.02(9) & 41(2)   \\ 
(5) & 2.5 & 16.8 & 0.4 & $30-50$   & $110-130$  & $3.5 - 4.8 $ & $8.3 - 10.5 $  & 18 & 0.11(5)  & 1.0(1)  & 29(4)   \\ 
(6) & 2.5 & 11.7 & 0.8 & $30-50$   & $110-130$  & $3.3 - 5.8 $ & $10.3 - 14.9 $ & 18 & 0.11(2)  & 0.99(4) & 24(1)   \\ 
(7) & 2.5 & 16.8 & 0.8 & $30-50$   & $110-130$  & $3.9 - 5.9 $ & $10.5 - 13.1 $ & 21 & 0.31(2)  & 0.75(2) & 16(1)   \\ 
(8)& 4.0 & 11.7 & 0.0 & $14-18$   & $30-34$    & $2.8 - 4.1 $ & $6.5 - 7.6 $   & 15 & 0.13(4)  & 1.15(8) & 13(1)   \\ 
(9)& 4.0 & 16.8 & 0.0 & $14-18$   & $30-34$    & $4.1 - 4.4 $ & $7.2 - 7.7 $   & 15 & 0.31(9)  & 1.0(1)  & 9(1)    \\ 
(10)& 4.0 & 11.7 & 0.4 & $14-18$   & $30-34$    & $5.5 - 6.3 $ & $7.9 - 8.7 $   & 15 & 1.2(2)   & 0.58(4) & 4(0)    \\ 
(11)& 4.0 & 16.8 & 0.4 & $14-18$   & $30-34$    & $5.8 - 7.3 $ & $10.2 - 11.9 $ & 15 & 1.0(2)   & 0.70(7) & 4(1)    \\ 
(12)& 4.0 & 11.7 & 0.8 & $14-18$   & $30-34$    & $6.6 - 8.1 $ & $10.9 - 13.2 $ & 15 & 0.5(2)   & 0.9(1)  & 6(1)    \\ 
(13)& 4.0 & 16.8 & 0.8 & $14-18$   & $30-34$    & $6.9 - 9.1 $ & $11.9 - 12.7 $ & 15 & 1.7(4)   & 0.58(7) & 2(1)    \\ 
\hline
\end{tabular}
\label{tab:qpkr_tab1}
\end{table}

\end{center}

For a particular dataset, we perform several power law fits, one for each pair of starting and ending points. We then determine values for $E_0$ and $\alpha$ for that dataset from the averages of all the values returned by the fits. We choose our ranges of start and end points carefully, in order to guard against our lack of precise knowledge of how the delocalization starts and also from systematic effects specific to our experiment. The starting ranges are chosen to be after some delocalization is clearly observable in the data. The ending points are chosen such that they are below the experimental timescale of atom loss due to finite trap depth \cite{seet22}. 

\begin{center}

\begin{table}[b!]
\caption{\bf{Ranges of pulse numbers used for fitting the data in Fig. \ref{fig:qpkr_figS2}, corresponding energy ranges, number of fits, and extracted values.}}
\begin{tabular}{|c|c|c|c|c|c|c|c|c|c|c|c|c|} 
\hline
Dataset  & $d$ & $K$ & $gn_{\rm 1D}$ & $\varepsilon$ & Beginning $n_p$ & Ending $n_p$  & Beginning $\Ez/\Erec$ & Ending $\Ez/\Erec$ & Number of fits & $E_0$ & $\alpha$ & $n_p^*$ \\
\hline
(1) & 1 & 2.5 & 11.7 & 0.4 & $30-50$   & $110-130$  & $1.5 - 2.2 $ & $3.9 - 4.5 $   & 17 & 0.15(6)  & 0.66(9) & 79(1)   \\
(2) & 1 & 2.5 & 16.8 & 0.4 & $30-50$   & $110-130$  & $2.9 - 4.5 $ & $8.2 - 10.4$   & 18 & 0.27(4)  & 0.71(3) & 22(1)   \\ 
(3) & 2 & 2.5 & 11.7 & 0.4 & $30-50$   & $110-130$  & $2.1 - 3.6 $ & $6.8 - 9.0 $   & 18 & 0.07(2)  & 0.97(7) & 41(2)   \\ 
(4) & 2 & 2.5 & 16.8 & 0.4 & $30-50$   & $110-130$  & $2.9 - 3.6 $ & $7.7 - 8.7 $   & 18 & 0.14(5)  & 0.86(8) & 31(3)   \\ 
(5) & 3 & 2.5 & 11.7 & 0.4 & $30-50$   & $110-130$  & $2.4 - 3.6 $ & $7.6 - 8.7 $   & 18 & 0.06(2)  & 1.02(9) & 41(2)   \\ 
(6) & 3 & 2.5 & 16.8 & 0.4 & $30-50$   & $110-130$  & $3.5 - 4.8 $ & $8.3 - 10.5$   & 18 & 0.11(5)  & 1.0(1)  & 29(4)   \\ 
(7)& 4 & 2.5 & 11.7 & 0.4 & $30-50$   & $110-130$  & $1.2 - 2.4 $ & $3.5 - 5.1 $   & 18 & 0.11(4)  & 0.78(7) & 60(2)   \\ 
(8)& 4 & 2.5 & 16.8 & 0.4 & $30-50$   & $110-130$  & $2.9 - 3.9 $ & $6.8 - 7.7 $   & 18 & 0.20(7)  & 0.76(8) & 31(3)   \\ 
\hline
\end{tabular}
\label{tab:qpkr_tab2}
\end{table}

\end{center}

\begin{figure}[t!]
    \center
    \includegraphics[width=1.0\textwidth]{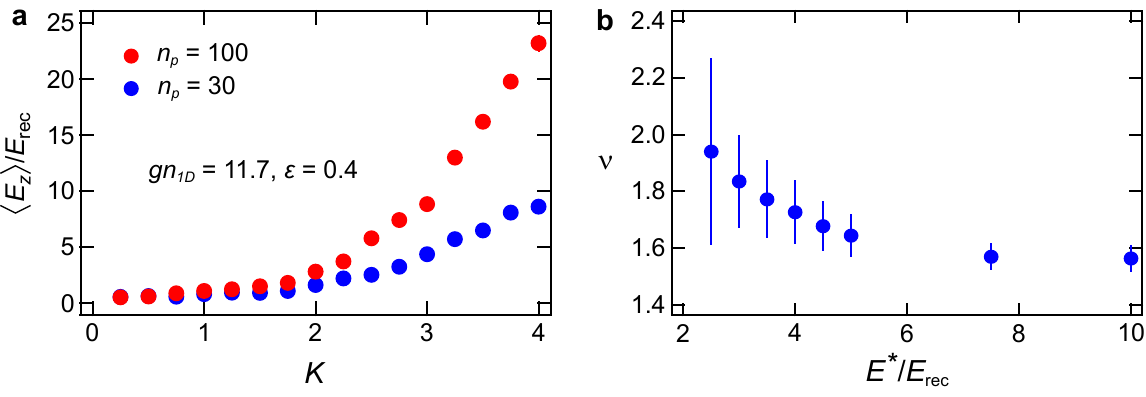}
    \caption{Two-point analysis for divergence of $n_p^*$ corresponding to Fig. 2(e) in main paper. (a) The energy of the system at $n_p=30$ and $100$ with $\gn=11.7$, $\varepsilon=0.4$, pulse parameters $t_p=4 \,\mu$s and $T=105\,\mu$s, at various kick strengths $K$. The delocalization onset time $n_p^*$ is calculated from the energy values at the two different pulse numbers. (b) Fitted critical exponent $\nu$ for different values of $E^*$.}
    \label{fig:qpkr_figS3}
\end{figure}

\begin{figure}[h!]
    \center
    \includegraphics[width=1.0\textwidth]{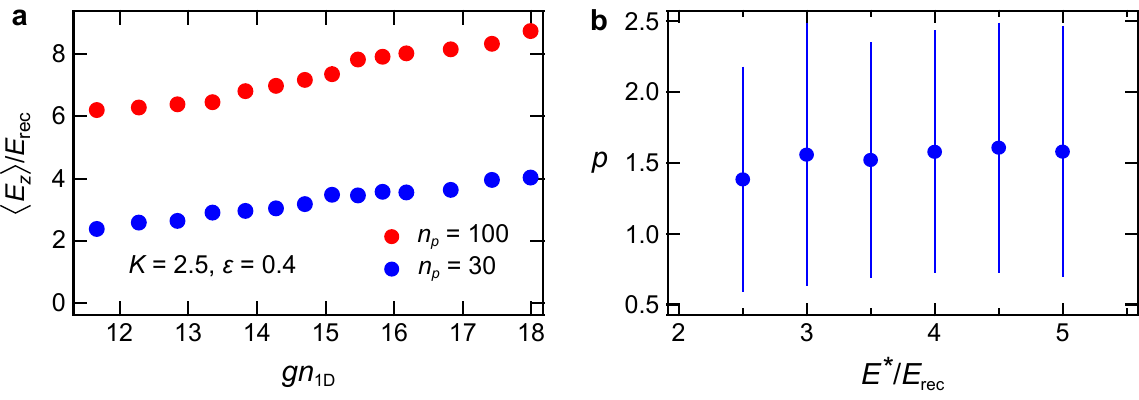}
    \caption{Two-point analysis for the variation of $n_p^*$ corresponding to Fig. 4(b) in main paper. (a) The energy of the system at $n_p=30$ and $100$ as $\gn$ is varied, keeping $K=2.5$, $\varepsilon=0.4$, pulse parameters $t_p=4 \,\mu$s and $T=105\,\mu$s. The delocalization onset time $n_p^*$ is calculated from the energy values at the two different pulse numbers. (b) Fitted decay exponent $p$ for different values of $E^*$.}
    \label{fig:qpkr_figS4}
\end{figure}

\noindent {\it Two point Analysis.---} For a power-law energy evolution, the delocalization parameters can also be extracted by measurements at two different pulse numbers. The data shown in Figs. 2(e) and 4(b) of the main paper were obtained using this two point analysis method which we now describe. We measured the energy of the system at pulse numbers $n_p=30$ and $n_p=100$, as shown in Figs. \ref{fig:qpkr_figS3} and \ref{fig:qpkr_figS4}, and then calculated the $n_p^*$ using the method outlined below. 

Assuming the delocalization dynamics follow a power-law behavior such that $\langle E_z \rangle =E_0 n_p^\alpha$, the average energy of the system at the two different pulse numbers $\langle E_z(30) \rangle$ and $\langle E_z(100) \rangle$ can then be expressed as
\begin{align}
    \langle E_z(30) \rangle &= E_0 \cdot 30^{\alpha} \label{eq:energyatpulse30} \\
    \langle E_z(100) \rangle &= E_0 \cdot 100^{\alpha}
    \label{eq:energyatpulse100}
\end{align}
Solving simultaneous equations (\ref{eq:energyatpulse100}) and (\ref{eq:energyatpulse30}) for $\alpha$ and $E_0$, we obtain
\begin{equation}
    \alpha= \frac{\ln{(\langle E_z(100) \rangle/ \langle E_z(30) \rangle)}}{\ln{(100/30)}} \, .
    \label{eq:alpha}
\end{equation}
and
\begin{equation}
    \ln(E_0) = \frac{\ln (100) \ln (\langle E_z(30) \rangle) - \ln(30) \ln(\langle E_z(100) \rangle) }{\ln{(100/30)}} \, .
    \label{eq:prefactor}
\end{equation}
Using Eq. (\ref{eq:alpha}) and Eq. (\ref{eq:prefactor}), we then solve for the delocalization onset time $n_p^*$, which is defined to be the pulse number when the energy of the system reaches $2.5\Erec$, i.e. $\langle E_z \rangle = E_0 (n_p^*)^\alpha = 2.5\Erec$. The equivalence of this method of determining $n_p^*$ to that of a power law fit to time evolution data comprising several $n_p^*$ is also demonstrated numerically in Fig. \ref{fig:qpkr_figS6}.

In Fig. \ref{fig:qpkr_figS3}(a) we show the data used to produce the plot in Fig 2(e) of the main paper where $E^*=2.5 E_{\rm rec}$ is used to evaluate $n_p^*$. Near the MIT phase boundary, $\langle E_z(100) \rangle$ approaches $\langle E_z(30) \rangle$ leading to a divergence of $n_p^*$ as well as an increase in the extracted error bar (see Fig 2(e) of the main paper).

The variation of $n_p^*$ with $K$ can be fit with a function of the form $1/|K-K_c|^{\nu}$ to extract a critical exponent $\nu$. The variation of the extracted $\nu$ with choice of $E^*$ is shown in Fig. \ref{fig:qpkr_figS3}(b). Even though a weak downward trend can be seen, $\nu$ is approximately independent of choice of $E^*$ within error bars.

In Fig. \ref{fig:qpkr_figS4}(a) we show the data used to produce the plot in Fig 4(b) of the main paper where $E^*=2.5 E_{\rm rec}$ is used to evaluate $n_p^*$. The variation of $n_p^*$ with $\gn$ can be fit with a function of the form $1/|\gn-gn_{\rm 1D,c}|^{p}$ to extract an exponent $p$ characterizing the interaction dependence. The variation of the extracted $p$ with choice of $E^*$ is shown in Fig. \ref{fig:qpkr_figS4}(b), and is independent of choice of $E^*$ within error bars.

\section{Theoretical Description}

{\it Mapping to the $d$-dimensional Anderson model.---} As discussed in the main text, the $d$-dimensional Anderson model can be realized in the non-interacting quasi-periodic kicked rotor (QPKR) by modulating the amplitude of kick pulses with $d-1$ incommensurate frequencies as
\begin{equation}
i\kbar \partial_\tau \Phi(\theta,\tau) =  \left[ -\frac{\kbar^2}{2} \partial^2_\theta  - K \left(1 +  \varepsilon \prod_{j=2}^{d} \cos \omega_j\tau\right) \cos \theta \sum_{n_p} \delta(\tau-n_p)  \right]  \Phi(\theta,\tau).
\label{eq:QPKR}
\end{equation}
Note that Eq. (1) of the main text corresponds to $d=3$ and includes a non-linear interacting term and a harmonic trap potential. Here we follow the approach of \cite{casati89} to map the non-interacting QPKR to the $d$-dimensional Anderson model. First, we show that the dynamics of the QPKR is identical to that of a $d$-dimensional periodically kicked rotor (dDPKR), described by the Schr{\"o}dinger equation
\begin{equation}
\begin{split}
    i\kbar \partial_\tau \Psi(\vec{\theta},\tau) = & \bigg[ -\frac{\kbar^2}{2} \partial^2_{\theta_1} - \sum_{j=2}^d i\kbar\omega_j\partial_{\theta_j}  - K \left(1 +  \varepsilon \prod_{j=2}^{d} \cos \theta_j\right) \cos \theta_1 \sum_{n_p} \delta(\tau-n_p)  \bigg]  \Psi(\vec{\theta},\tau)
\end{split}
\label{dDPKR}
\end{equation}
with the initial condition 
\begin{equation}
    \Psi(\vec{\theta},\tau=0)=\Phi'(\theta_1,\tau=0)\prod_{j=2}^d \delta(\theta_j),
    \label{eq:initial}
\end{equation}
where $\vec{\theta}=(\theta_1,\theta_2,\cdots,\theta_d)$.
Eq.~(\ref{eq:initial}) indicates that the initial state is maximally localized in the additional $d-1$ dimensions at $\theta_{j\geq 2}=0$, while $\Phi'(\theta_1,\tau=0)$ can be an arbitrary wavefunciton. Furthermore, the kinetic terms in Eq. (\ref{dDPKR}) feature a linear dispersion $-i\kbar\omega_j\partial_{\theta_j}$ (for $j\geq2$) where $-i\kbar\partial_{\theta_j}$ is the momentum operator.
Due to the periodic kick, the time evolution of $\Psi$ is governed by the Floquet operator
\begin{equation}
    U=e^{iK \left(1 +  \varepsilon \prod_{j=2}^{d} \cos \theta_j\right) \cos \theta_1/\kbar} e^{-i\left(-\frac{\kbar^2}{2} \partial^2_{\theta_1} - \sum_{j=2}^d i\kbar\omega_i\partial_{\theta_j}\right)/\kbar},
\end{equation}
from a stroboscopic point of view. Note that the kick period is chosen as unity. After $\tau$ kicks, the wavefunciton becomes
\begin{equation}
\begin{split}
    \Psi(\theta_1,\cdots,\theta_d,\tau) & = U^\tau \Psi(\theta_1,\cdots,\theta_d,0) \\
    & = \Phi'(\theta_1,\tau) \prod_{j=2}^d \delta(\theta_j - \omega_j\tau), 
\end{split}
\end{equation}
where  
\begin{equation}
    \Phi'(\theta_1,\tau) = \prod_{\tau'=1}^\tau e^{iK \left(1 +  \varepsilon \prod_{j=2}^{d} \cos \omega_j\tau'\right) \cos \theta_1/\kbar} e^{i\frac{\kbar^2}{2} \partial^2_{\theta_1} /\kbar}\Phi'(\theta_1,0). 
\end{equation}
Comparing to Eq. (\ref{eq:QPKR}), it shows that the time evolution of $\Phi'(\theta_1,\tau)$ in the dDPKR follows exactly the same Schr{\"o}dinger equation as that of $\Phi(\theta,\tau)$ in the QPKR. Here we use the fact that $e^{-i(-i\omega_j\partial_{\theta_j})}$ is nothing more than the translation operator and hence $e^{-i(-i\omega_j\partial_{\theta_j})}\delta(\theta_j)=\delta(\theta-\omega_j)$. Because the $d$-dimensional wavefunction $\Psi(\vec{\theta},\tau)$ remains maximally localized in $\theta_{j\geq2}$, the change of kinetic energy due to kicks merely comes from the time evolution of $\Phi'(\theta_1,\tau)$ and is identical to that of the QPKR when $\Phi'(\theta_1,0)=\Phi(\theta_1,0)$. Therefore, the QPKR and dDPKR are dynamically equivalent.

The advantage of introducing the dDPKR is that it is a Floquet system and hence the wavefunction of the system can be expressed as $\Psi(\vec{\theta},\tau)=e^{-i\omega\tau}\psi(\vec{\theta},\tau)$, where $\omega$ is the quasienergy and $\psi(\vec{\theta},\tau)=\psi(\vec{\theta},\tau+1)$. Then it can be mapped to the $d$-dimensional Anderson model in momentum space. Considering the wavefunction just after and before the kick, which are denoted as $\psi_\pm(\vec{\theta})$, we have the relation
\begin{equation}
    \psi_+(\vec{\theta}) = e^{iK \left(1 +  \varepsilon \prod_{j=2}^{d} \cos \theta_j\right) \cos \theta_1/\kbar }  \psi_-(\vec{\theta}).
\end{equation}
On the other hand, the free evolution between two adjacent kicks yields
\begin{equation}
    \psi_-(\vec{\theta}) = e^{i\left(\omega+\frac{\kbar^2}{2} \partial^2_{\theta_1} + \sum_{j=2}^d i\kbar\omega_i\partial_{\theta_j}\right)/\kbar} \psi_+(\vec{\theta}).
\end{equation}
To map to the momentum space, we perform the Fourier transformation $\psi_\pm(\vec{\theta}) = \frac{1}{\sqrt{\Omega}}\int d \vec{\theta}e^{i\vec{k}\cdot\vec{\theta}} \psi_{\pm,\vec{k}}$ where $\Omega$ is the system volume and $\vec{k}$ is the $d$-dimensional momentum. From the two equations above, we derive the $d$-dimensional Anderson model in momentum space as
\begin{equation}
    V_{\vec{k}}\psi_{\vec{k}} + \sum_{\vec{k}'\neq 0} K_{\vec{k}'}\psi_{\vec{k}+\vec{k}'} = \varepsilon \psi_{\vec{k}},
\end{equation}
with the on-site disorder $V_{\vec{k}} = \tan\left(\omega/2\kbar-\kbar k_1^2/4-\sum_{j=2}^d\omega_j k_j/2\right)$, hopping rates $K_{\vec{k}}=\frac{1}{\Omega}\int d\theta e^{i\vec{k}\cdot\vec{\theta}}\tan\big[ \frac{K}{2\kbar}\big(1+\varepsilon\prod_{j=2}^d\cos\theta_j\big) \cos\theta_1 \big]$, $\psi_{\vec{k}}=(\psi_{-,\vec{k}}+\psi_{+,\vec{k}})/2$, and $\varepsilon = -K_{\vec{0}}$.

\section{Numerical Simulation}

In numerical simulations of all figures, we take into account the effect of the Gaussian shaped trap and atom loss from its finite depth in actual experiment. In Eq. (1) in the main text, the trap term is taken as harmonic and infinitely deep for the simplicity of the presentation. Furthermore, we scale the trap in order to maintain the same frequency for different $s_{\bot}$ to match the constant axial frequency. In the experiment, $s_{\bot}$ is changed in order to obtain different interaction strength $g$, which, however, also change $\omega_z$. To overcome this experimentally, we kept the ODT on to maintain an approximately constant axial frequency. The scaled trap ensures consistent trapping terms to capture the effects from only changing the interaction term. \\

\begin{figure}[h!]

\begin{tabular}{|c|c|c|c|} 
\hline
        & Energy Evolution and Fitting &$n_p^*$ &Exponent $\alpha$\\
\hline
\rotatebox{90}{\qquad $gn1D=11.7$} & \includegraphics[width=0.25\textwidth]{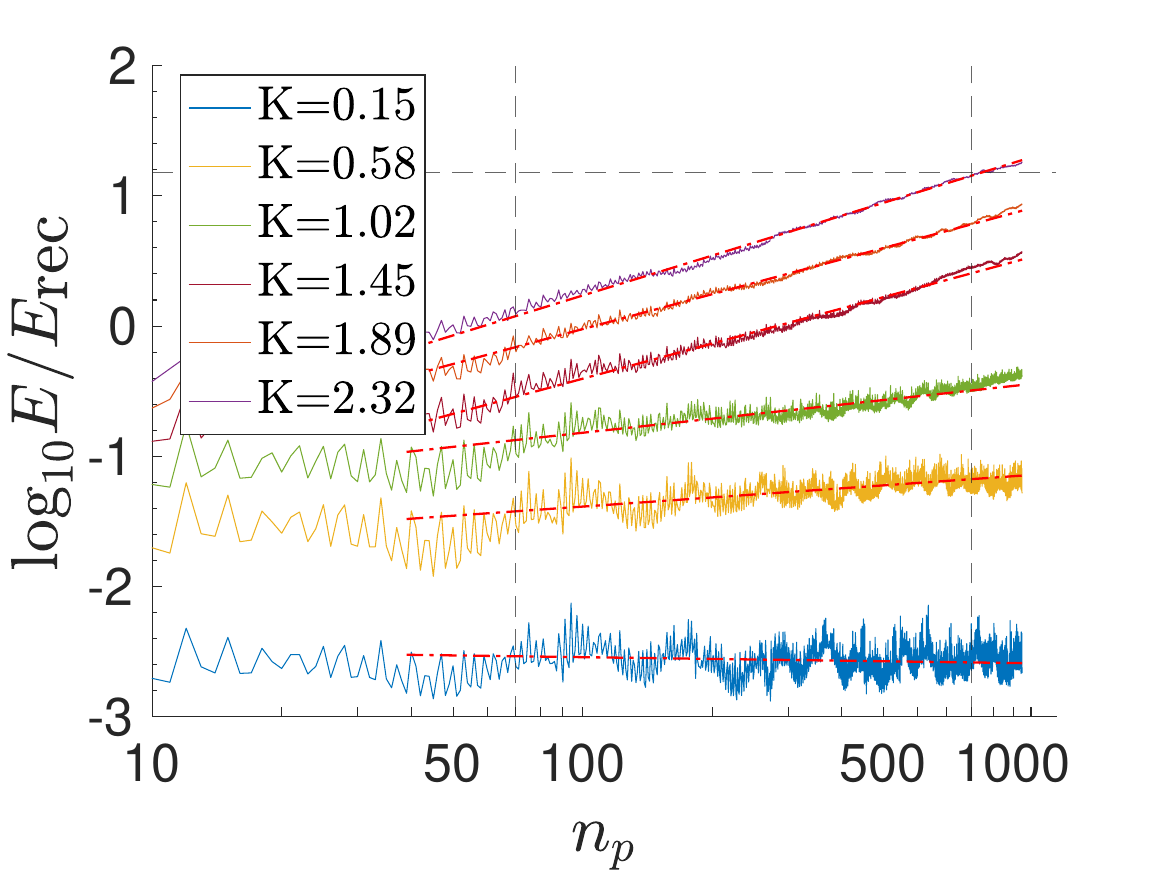} & \includegraphics[width=0.3\textwidth]{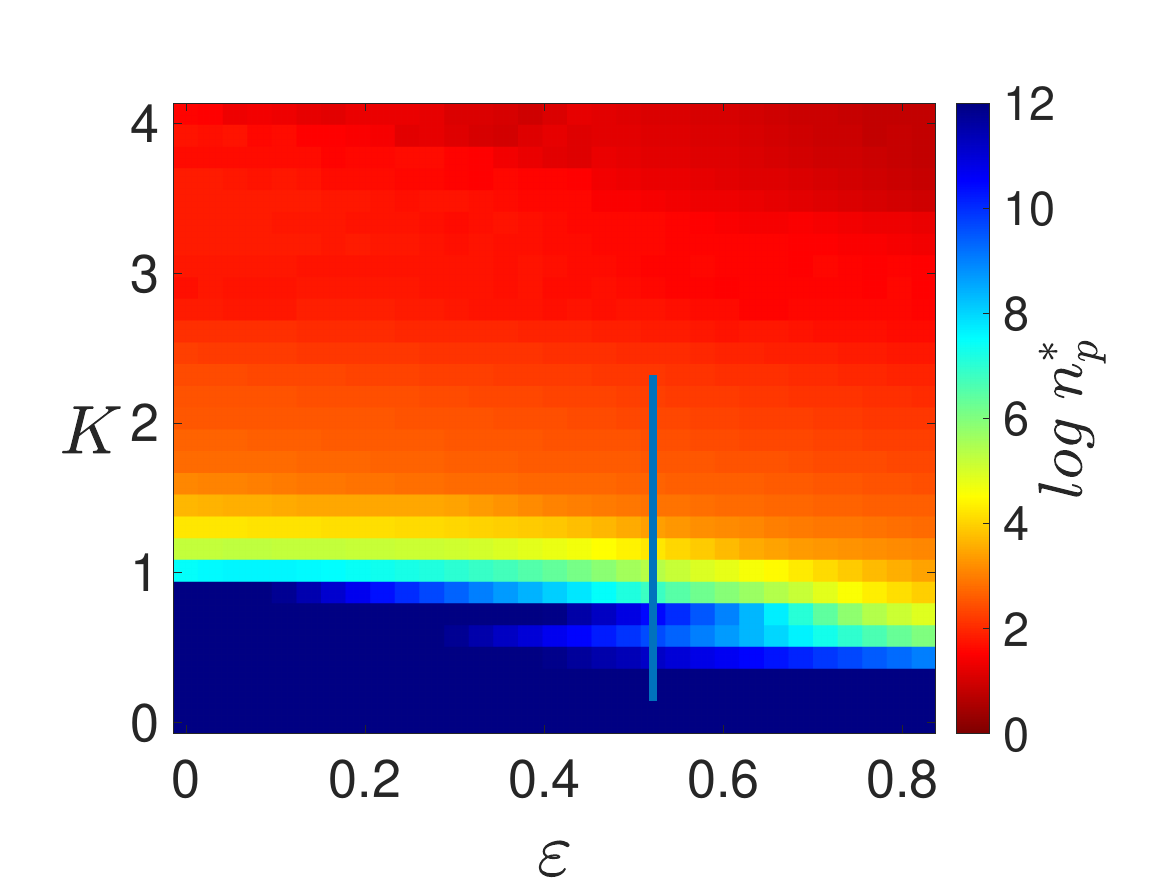} & \includegraphics[width=0.3\textwidth]{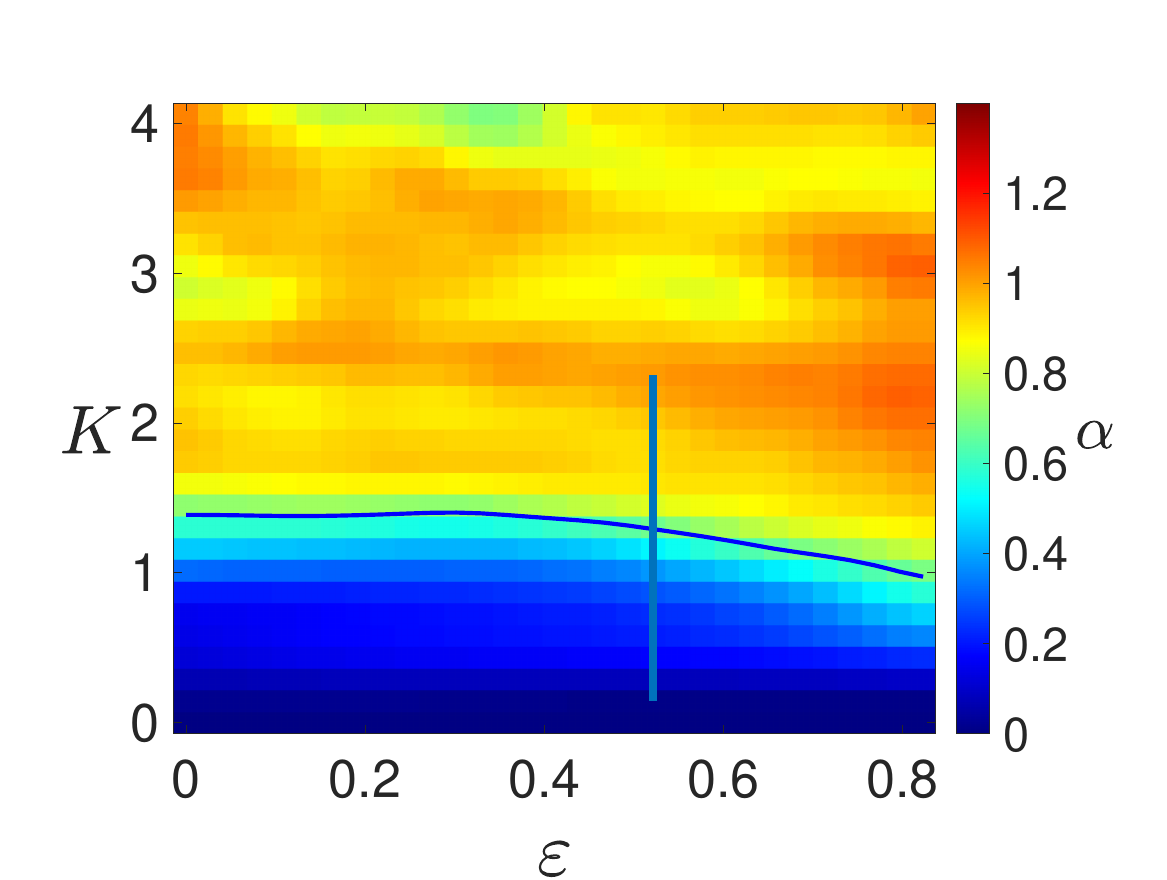}   \\ 
\hline
\rotatebox{90}{\qquad $gn1D=16.8$} & \includegraphics[width=0.25\textwidth]{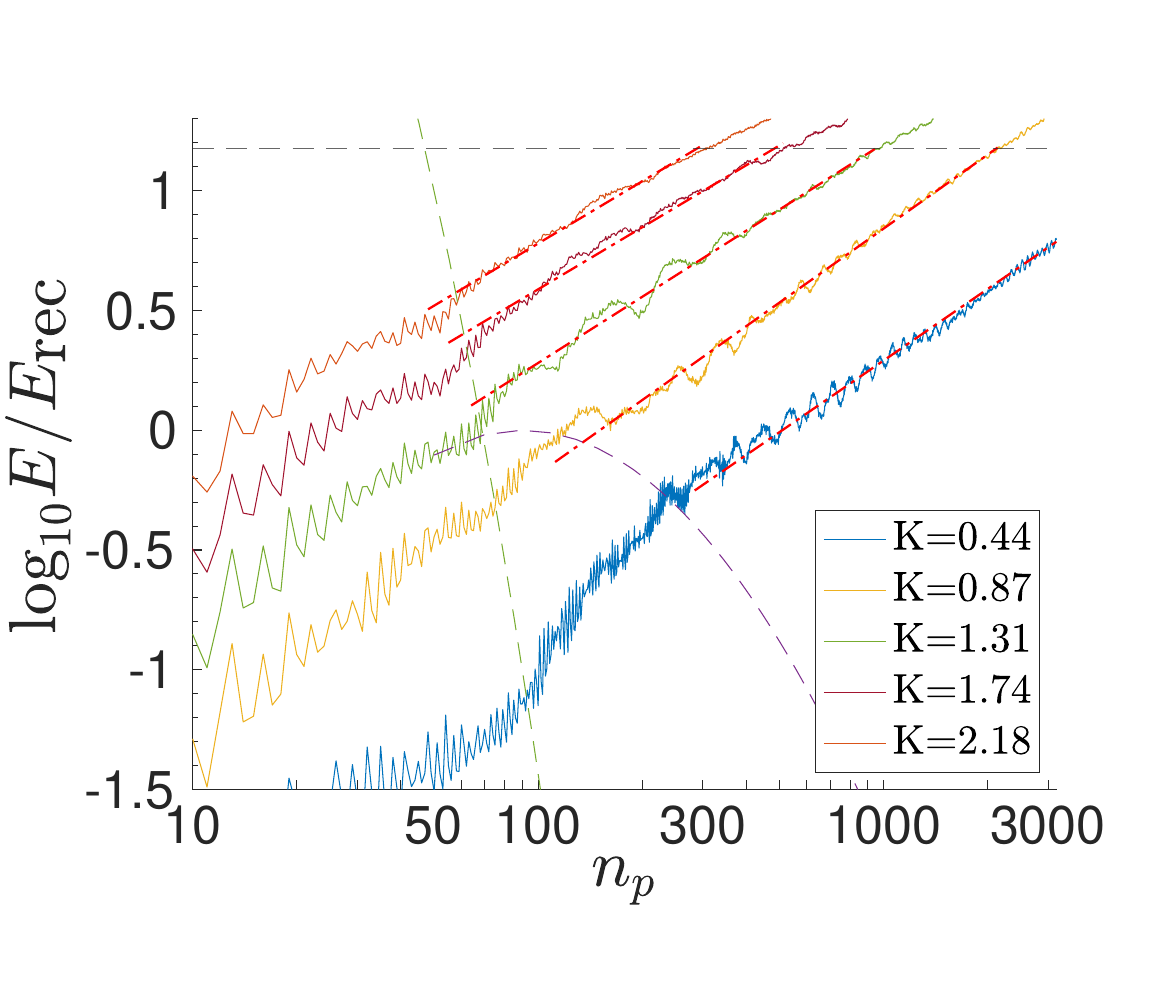} &\includegraphics[width=0.3\textwidth]{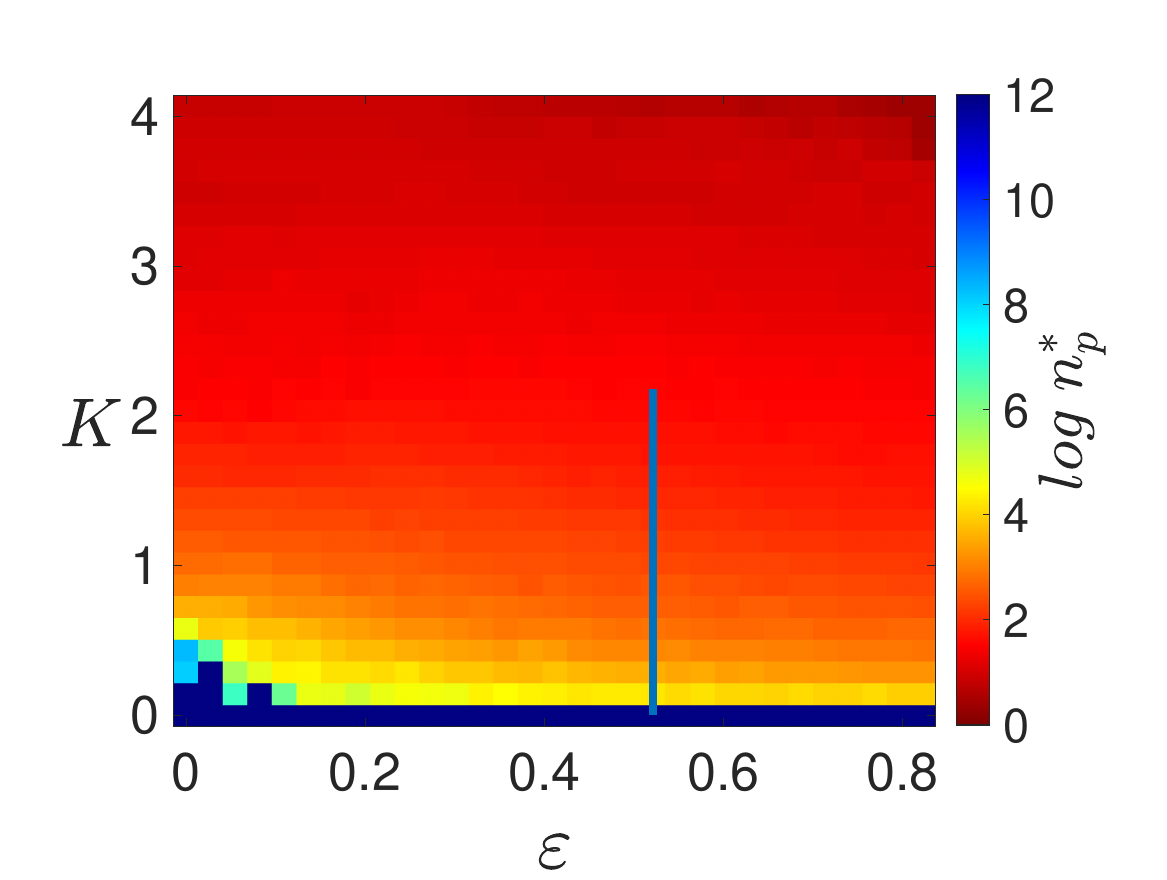} & \includegraphics[width=0.3\textwidth]{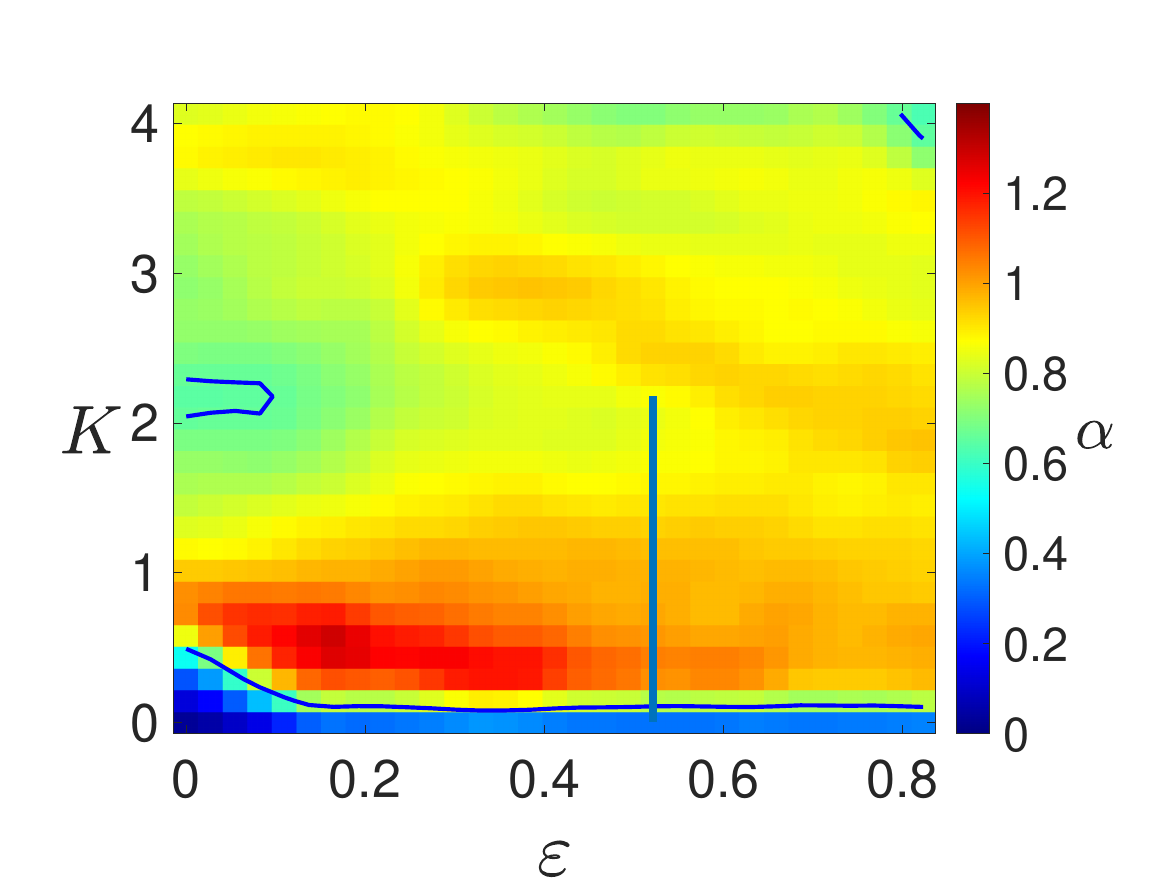}     \\ 
\hline
\end{tabular}
\\
\caption{Phase Diagrams for $\gn=$11.7 and 16.8. Left panels show the energy evolution curves for various $K$ and $\varepsilon = 0.52$, together with the corresponding power law fits (linear on log-log plot). Middle panels show the phase diagrams for $n_p^*$. Right panels show the phase diagrams for the exponent $\alpha$.}
\label{fig:qpkr_figS5}
\end{figure}

The dimensionless GP equation in numerical simulation of a single tube is:
\begin{align*}
i\frac{\partial}{\partial \tau}&\tilde{\psi}(\tilde{z},\tau)=\Bigl( \tilde{p}^2+ V(\tilde{z})+\tilde{g}\left|\tilde{\psi}(\tilde{z},\tau)\right|^2\\
&+\frac{s_z}{2}\cos 2\tilde{z}\left(1+\varepsilon\cos(\tilde{\omega}_2 \tau)\cos(\tilde{\omega}_3 \tau)\right)\sum_n\delta(\tau-n\tilde{T})\Bigr)\tilde{\psi}(\tilde{z},\tau)
\end{align*}
 where $\tilde{p} = p/\hbar k_L$, $\tilde{z} = k_L z$, $\tau = E_R t/\hbar$, $\tilde{g} = k_L g/E_R $ and $\tilde{\psi}(\tilde{z}) =\psi(\tilde{z})/\sqrt{k_L} $ . $V(\tilde{z})$ is the Gaussian shaped trap representing optical dipole potential:\\
 \begin{equation}
     V(\tilde{z}) = -s_\perp \left(e^{-\frac{2 \tilde{z} ^2}{w_x^2}} +e^{-\frac{2 \tilde{z} ^2}{w_y^2}} \right),
 \end{equation}
$\omega_x $ and $ \omega_y$ are the lattice beam waists while $s_\perp$ and $s_z$ are laser strengths introduced in the main text. The wavefunction is normalized as $\sum\left|\tilde{\psi}(\tilde{z})\right|^2 \Delta \tilde{z} =1$. The atom loss is considered by lowering $\tilde{g}$ since interaction is proportional to atom number in mean field approximation. Notice that the Hamiltonian above can be separated into a position-diagonal part and a momentum-diagonal part. Hence for every time step (in most simulations it is $10^{\text{-2}}\mu s$), we implement Suzuki-Trotter expansion to the time evolution operator. The time evolution is then performed in every tube and final energies in plots are obtained by averaging over all tubes. Atom numbers of tubes are approximated by assuming the condensate follows a Thomas-Fermi distribution in the trap before the kicks are applied. Details about tube averaging are in the supplement of our previous work \cite{seet22}.\\

\noindent {\it Phase Diagrams.---} For the phase diagrams in Fig. 1 of the main paper, we simulated the system in a single tube with an average atom number $N_{\rm atom}=300$. For a given $K$ and $\varepsilon$, the system evolves from the same initial state, obtained from imaginary time evolution (iterating time: $2\times 10^{5}$), but with different kicking parameters. In other words, every point in the phase diagram corresponds to a different energy evolution curve. 

\vspace{-0.4cm}
\begin{figure}[h]
%\subfigure[]{\label{fig:Onset_fitting}\includegraphics[width=0.4\textwidth]{FigureS6a.pdf}}
%\subfigure[]{\label{fig:Onset_different_definition}\includegraphics[width=0.4\textwidth]{FigureS6b.pdf}}
\includegraphics[width=0.8\textwidth]{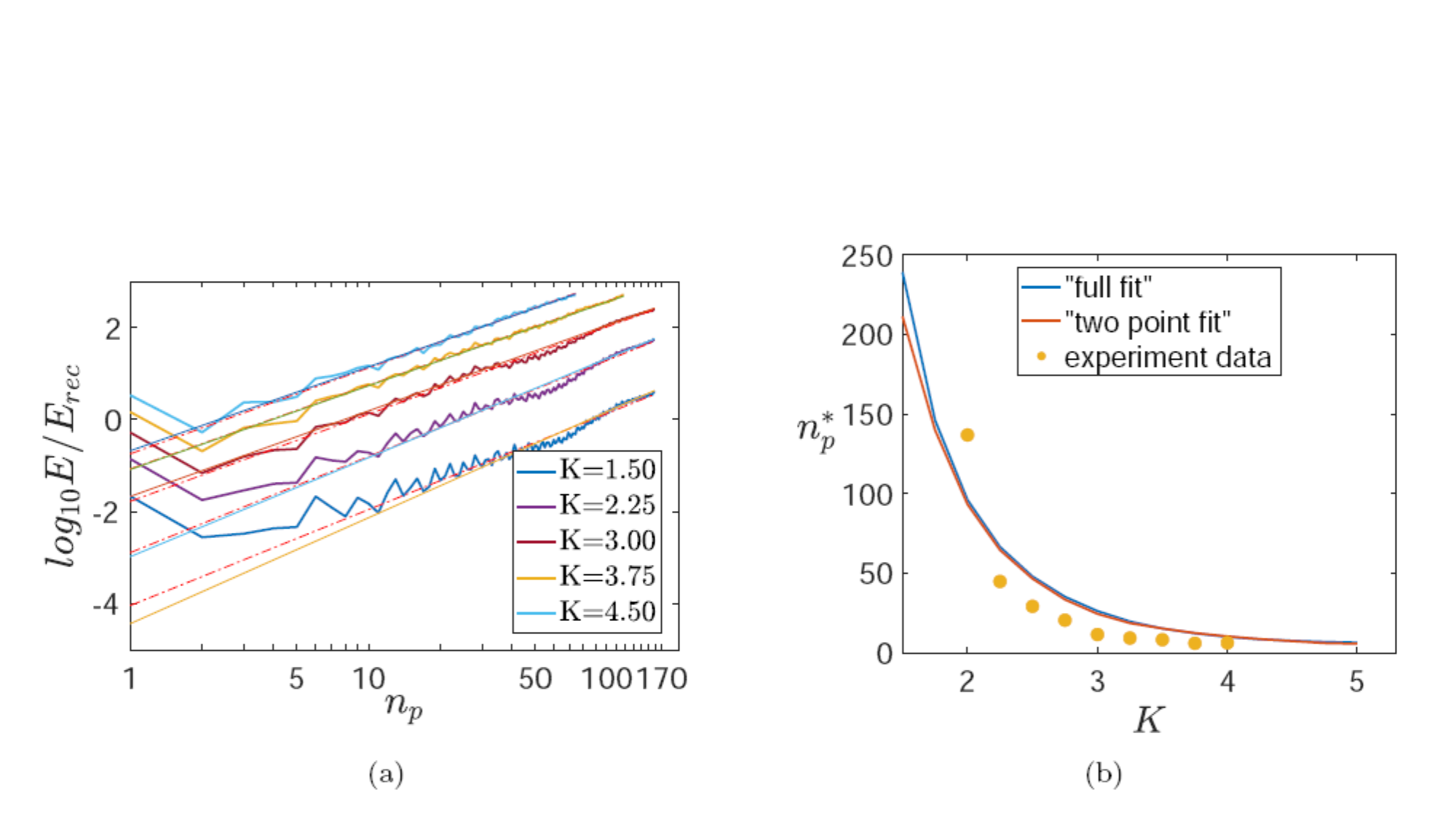}
\caption{Comparison of the two methods to determine onset time $n_p^*$. In (a), dashed lines are the ``full fit" lines for the energy curves, corresponding to the same fitting method used to generate the phase diagrams (Fig. 1 in main text and Fig. \ref{fig:qpkr_figS5}), and solid straight lines are the ``two point fit" lines. The parameters in this figure correspond to Fig. 2 in the main text. The solid lines in (b) show $n_p^*$ extracted from the``two point fit" method and the ``full fit" method, from the same energy evolution shown in (a). As can be seen, these two methods only show very small differences. The experimental data corresponding to Fig. 2(e) in the main text are also shown. }
\label{fig:qpkr_figS6}
\end{figure}

Examples of numerically simulated energy evolution curves and the phase diagrams generated for $\gn=11.7$ and 16.8 are shown in Fig. \ref{fig:qpkr_figS5}. Each curve on the left panel corresponds to a different point positioned along the solid lines in the phase diagrams in the middle and right panels. As shown in the left panels, the system initially fluctuates. We choose our fitting region in the subsequent linear part (in the log-log plot), allowing the ranges to be different for different curves. By observing the obvious pattern of the linear evolution, the starting and end points of the fitting region are obtained from the intersection of the dashed lines and our energy curves. The end points are also limited to $15 E_{\rm rec}$ (horizontal dashed line) to avoid effects from particle loss from the finite trap depth ($60\sim 180 \,E_\text{rec}$). Finally, the onset time $n_p^*$ of each curve is estimated as the pulse number when the fitting line intersects with the energy $E^*= 2.5\,E_\text{rec}$. 

The phase diagrams shown in the middle and right panels for the onset time $n_p^*$ and exponent $\alpha$ respectively both clearly show a sharp change in behavior across a phase boundary indicative of a many-body MIT. For each interaction strength, even though the phase boundaries for $n_p^*$ and $\alpha$ are in fair agreement, the two phase diagrams exhibit considerable difference within the delocalized metallic phase. This is a consequence of the inhomogeneity in the system arising from the harmonic trap where single particle effects lead to additional features in the $\alpha$ phase diagram over longer times. These effects are suppressed in the $n_p^*$ phase diagrams for appropriately chosen $E^*=2.5 E_{\rm rec}$ which is sufficiently larger than the initial energy, yet small enough for a delocalized system to reach on a timescale shorter than the axial oscillation period (about 150 pulses). 

In Fig. \ref{fig:qpkr_figS6}, we compare the two methods to determine $n_p^*$ that we have discussed - extraction from fitting a power law to time evolution data and the ``two-point" analysis. The comparison is made for the case of the numerical simulation for Fig. 2(e) of the main paper, where a divergence of $n_p^*$ is observed as an indication of a many-body Anderson MIT. The onset of delocalization is defined as $n_p^*=(2.5E_{\rm rec}/E_0)^{\frac{1}{\alpha}}$. Fig. \ref{fig:qpkr_figS6} shows that the difference between these two types of estimates (two solid lines in Fig. \ref{fig:qpkr_figS6}(b)) is negligible. \\

\end{document}